\documentclass[]{spie}  

\usepackage{amsmath,amsfonts,amssymb}
\usepackage{graphicx}
\usepackage[colorlinks=true, allcolors=blue]{hyperref}

\usepackage{mathtools}
\usepackage{mathrsfs}
\makeatletter
\newcommand{\ostar}{\mathbin{\mathpalette\make@circled\star}}
\newcommand{\make@circled}[2]{%
  \ooalign{$\m@th#1\smallbigcirc{#1}$\cr\hidewidth$\m@th#1#2$\hidewidth\cr}%
}
\newcommand{\smallbigcirc}[1]{%
  \vcenter{\hbox{\scalebox{0.77778}{$\m@th#1\bigcirc$}}}%
}

\usepackage{array}
\usepackage{makecell}
\newcolumntype{C}[1]{>{\centering\arraybackslash}p{#1}} 
\newcolumntype{L}[1]{>{\raggedright\arraybackslash}m{#1}} 
\newcolumntype{M}[1]{>{\centering\arraybackslash}m{#1}} 
\newcolumntype{B}[1]{>{\centering\arraybackslash}b{#1}} 

\usepackage{graphicx}
\graphicspath{{./figures/}}

\usepackage{transparent}
\usepackage{xkeyval,xcolor}
\makeatletter
\newlength{\sfp@hseplen}\newlength{\sfp@vseplen}
\define@cmdkey{subfigpos}[sfp@]{pos}[ul]{}
\define@cmdkey{subfigpos}[sfp@]{font}[\small]{}
\define@cmdkey{subfigpos}[sfp@]{vsep}[0.75\baselineskip]{\setlength{\sfp@vseplen}{\sfp@vsep}}
\define@cmdkey{subfigpos}[sfp@]{hsep}[3.5pt]{\setlength{\sfp@hseplen}{\sfp@hsep}}
\newcommand{\subfigimg}[4][,]{%
        \setkeys{Gin,subfigpos}{pos,font,vsep,hsep,#1}
        \setbox1=\hbox{\includegraphics{#4}}
        \ifnum\pdfstrcmp{\sfp@pos}{ul}=0
                \leavevmode\rlap{\usebox1}
                \rlap{\hspace*{\sfp@hsep}\raisebox{\dimexpr\ht1-\sfp@vsep}{\transparent{#3}{\setlength{\fboxsep}{1pt}\colorbox{white}{%
\transparent{1}\sfp@font{#2}}}%
}}
                \phantom{\usebox1}
        \else\ifnum\pdfstrcmp{\sfp@pos}{ur}=0
                \leavevmode\usebox1
                \llap{\raisebox{\dimexpr\ht1-\sfp@vsep}{\sfp@font{#2}}\hspace*{\sfp@hsep}}
        \else\ifnum\pdfstrcmp{\sfp@pos}{lr}=0
                \leavevmode\usebox1
                \llap{\raisebox{\sfp@vsep}{\sfp@font{#2}}\hspace*{\sfp@hsep}}
        \else
                \leavevmode\rlap{\usebox1}
                \rlap{\hspace*{\sfp@hseplen}\raisebox{\sfp@vsep}{\sfp@font{#2}}}
                \phantom{\usebox1}
        \fi\fi\fi
}
\newcommand{\fontfig}[1]{\small$\!\!$\color{#1}\textbf}
\newcommand{\AspectRatio}[1]{\dimexpr 1pt * \wd#1 / \ht#1 \relax} 

\renewcommand{\refeq}[1]{Eq.~(\ref{#1})\xspace} 
\newcommand{\refeqs}[2]{Eqs.~(\ref{#1}) and (\ref{#2})\xspace}
\newcommand{\refeqss}[3]{Eqs.~(\ref{#1}), (\ref{#2}) and (\ref{#3})\xspace}
\newcommand{\refeqfull}[1]{Equation~(\ref{#1})\xspace} 

\newcommand{\reffig}[1]{Fig.~\ref{#1}\xspace}

\newcommand{\refsub}[1]{#1} 
\newcommand{\refsubfig}[2]{Fig.~\ref{#1}\refsub{#2}\xspace}
\newcommand{\refsubfigfull}[2]{Figure~\ref{#1}\refsub{#2}\xspace} 
\newcommand{\refsubfigs}[2]{Figs.~\ref{#1}\refsub{#2}\xspace}
\newcommand{\refsubfigsfull}[2]{Figures~\ref{#1}\refsub{#2}\xspace} 

\newcommand{\refpan}[1]{Panel~#1} 
\newcommand{\refpans}[1]{Panels~#1} 

\newcommand{\reftab}[1]{Table~\ref{#1}\xspace}
\newcommand{\reftabs}[2]{Tables~\ref{#1} and~\ref{#2}\xspace}

\newcommand{\refsec}[1]{Sec.~\ref{#1}\xspace} 

\newcommand{\citet}[1]{Ref.~\citenum{#1}\xspace}

\DeclarePairedDelimiterX{\paren}[1]{(}{)}{#1}
\newcommand{\Paren}[1]{\paren*{#1}}
\let\brace=\undefined 
\DeclarePairedDelimiterX{\brace}[1]{\{}{\}}{#1}
\newcommand{\Brace}[1]{\brace*{#1}}
\let\brack=\undefined 
\DeclarePairedDelimiterX{\brack}[1]{[}{]}{#1}
\newcommand{\Brack}[1]{\brack*{#1}}
\DeclarePairedDelimiterX{\bbrack}[1]{\llbracket}{\rrbracket}{#1}

\DeclarePairedDelimiterX{\abs}[1]{\rvert}{\lvert}{#1}     

\DeclarePairedDelimiterX{\norm}[1]{\lVert}{\rVert}{#1}    

\DeclarePairedDelimiterX{\avg}[1]{\langle}{\rangle}{#1}   

\DeclarePairedDelimiterX{\ceil}[1]{\lceil}{\rceil}{#1}     

\DeclarePairedDelimiterX{\floor}[1]{\lfloor}{\rfloor}{#1}  

\newcommand{\Tag}[1]{\text{#1}}                 

\newcommand{\strehl}{\rho_\Tag{Strehl}}
\newcommand{\secant}{\chi}
\newcommand{\vWind}{v_{0}}
\newcommand{\hWind}{\hTag{0}}
\newcommand{\dtel}{D_\Tag{tel}}
\newcommand{\dwfs}{D_\Tag{wfs}}
\newcommand{\dTag}[1]{D_\Tag{#1}}
\newcommand{\pixscale}{p_\Tag{pix}}
\newcommand{\pixscaleTag}[1]{p_\Tag{pix,#1}}

\newcommand{\sig}[1]{\sigma_\Tag{#1}}
\newcommand{\sigTwo}[1]{\sigma^2_\Tag{#1}}
\newcommand{\rFried}[1]{r_\Tag{#1}}

\newcommand{\lambdaTag}[1]{\lambda_\Tag{#1}}
\newcommand{\fwhmTag}[1]{\Xi_\Tag{#1}}
\newcommand{\thetaTag}[1]{\theta_\Tag{#1}}
\newcommand{\hTag}[1]{h_\Tag{#1}}
\newcommand{\fluxTag}[1]{\Phi_\Tag{#1}}

\newcommand{\alphaTag}[1]{\alpha_\Tag{#1}}
\newcommand{\betaTag}[1]{\beta_\Tag{#1}}
\newcommand{\nTag}[1]{n_\Tag{#1}}

\newcommand{\dact}{d_\Tag{act}}

\newcommand{\X}[1]{$#1\!\times\!#1$}

\usepackage{siunitx}
\newcommand{\percent}[1]{#1\,\%}     

\usepackage{xspace}
\newcommand{\aspro}{ASPRO\textcolor{green!65!black}{$_2$}\xspace}

\allowdisplaybreaks

\title{Simplified model(s) of the GRAVITY+ adaptive optics system(s) for performance prediction}

\author[a]{Anthony Berdeu}
\author[b]{Jean-Baptiste Le Bouquin}
\author[b,c]{Guillaume Mella}
\author[b,c]{Laurent Bourgès}
\author[b]{Jean-Philippe Berger}
\author[d]{Guillaume Bourdarot}
\author[a]{Thibaut Paumard}
\author[d]{Frank Eisenhauer}
\author[e]{Christian Straubmeier}
\author[f,g]{Paulo Garcia}
\author[h]{Sebastian Hönig}
\author[i]{Florentin Millour}
\author[j]{Laura Kreidberg}
\author[k]{Denis Defrère}
\author[l]{Ferréol Soulez}
\author[d]{Taro Shimizu}

\affil[a]{LESIA, Observatoire de Paris, Université PSL, Sorbonne Université, Université Paris Cité, CNRS, 5 place Jules Janssen, 92195 Meudon, France}
\affil[b]{Univ. Grenoble Alpes, CNRS, IPAG, 38000 Grenoble, France}
\affil[c]{Univ. Grenoble Alpes, CNRS, IRD, INRAE, Météo France, OSUG, 38000 Grenoble, France}
\affil[d]{Max Planck Institute for extraterrestrial Physics, Giessenbachstraße 1, 85748 Garching, Germany}
\affil[e]{1st Institute of Physics, University of Cologne, Zülpicher Straße 77, 50937 Cologne, Germany}
\affil[f]{CENTRA - Centro de Astrofísica e Gravitação, IST, Universidade de Lisboa, 1049-001 Lisboa, Portugal}
\affil[g]{Faculdade de Engenharia, Universidade do Porto, rua Dr. Roberto Frias, 4200-465 Porto, Portugal}
\affil[h]{School of Physics \& Astronomy, University of Southampton, Southampton, SO17 1BJ, UK}
\affil[i]{Université Côte d’Azur, Observatoire de la Côte d'Azur, CNRS, Laboratoire Lagrange, France}
\affil[j]{Max Planck Institute for Astronomy, Königstuhl 17, 69117 Heidelberg, Germany}
\affil[k]{Institute of Astronomy, KU Leuven, Celestijnenlaan 200D, B-3001, Leuven, Belgium}
\affil[l]{Univ Lyon, Univ Lyon1, Ens de Lyon, Centre de Recherche Astrophysique de Lyon, UMR 5574, F-69230, Saint-Genis-Laval, France}

\authorinfo{On the behalf of the GRAVITY+ Collaboration. Send correspondence to Anthony Berdeu: \href{mailto:anthony.berdeu@obspm.fr}{anthony.berdeu@obspm.fr}}

\begin{document} 
\maketitle

\begin{abstract}
In the context of the GRAVITY+ upgrade, the adaptive optics (AO) systems of the GRAVITY interferometer are undergoing a major lifting. The current CILAS deformable mirrors (DM, 90 actuators) will be replaced by ALPAO kilo-DMs (\X{43}, 1432 actuators). On top of the already existing \X{9} Shack-Hartmann wavefront sensors (SH-WFS) for infrared (IR) natural guide star (NGS), new \X{40} SH-WFSs for visible (VIS) NGS will be deployed. Lasers will also be installed on the four units of the Very Large Telescope to provide a laser guide star (LGS) option with \X{30} SH-WFSs and with the choice to either use the \X{9} IR-WFSs or \X{2} VIS-WFSs for low order sensing. Thus, four modes will be available for the GRAVITY+ AO system (GPAO): IR-NGS, IR-LGS, VIS-NGS and VIS-LGS. To prepare the instrument commissioning and help the observers to plan their observations, a tool is needed to predict the performances of the different modes and for different observing conditions (NGS magnitude, science object magnitude, turbulence conditions, ...). We developed models based on a Maréchal approximation to predict the Strehl ratio of the four GPAO modes in order to feed the already existing tool that simulates the GRAVITY performances. Waiting for commissioning data, our model was validated and calibrated using the TIPTOP toolbox, a Point Spread Function simulator based on the computation of Power Spectrum Densities. In this work, we present our models of the NGS modes of GPAO and their calibration with TIPTOP.
\end{abstract}


\section{INTRODUCTION}
\label{sec:intro}

GRAVITY+\cite{GRAVITYplus:22_messenger} is a combined upgrade of the GRAVITY instrument and of the Very Large Telescope Interferometer\cite{GRAVITY:17_VLTI} (VLTI) of the European Southern Observatory (ESO). This work focuses on the major update of the adaptive optics (AO) systems of the four Unit Telescopes (UTs) of the VLTI. The role of an AO system is to compensate for the atmospheric turbulence\cite{Roddier:99_AO_system}. To do so, it is composed by a wavefront sensor (WFS) whose measurements are analyzed by a real time computer (RTC) and converted into a set of commands sent to a deformable mirror (DM) that corrects the optical aberrations. This feedback loop must run faster than the turbulence temporal evolution (typically with a frequency ranging from several hundred Hertz to a kilo-Hertz).

\begin{figure}[h!] 
        \centering
        
        \newcommand{\PathFig}{Fig_ASPRO_GPAO_modes/}
        
        \includegraphics[width=\linewidth]{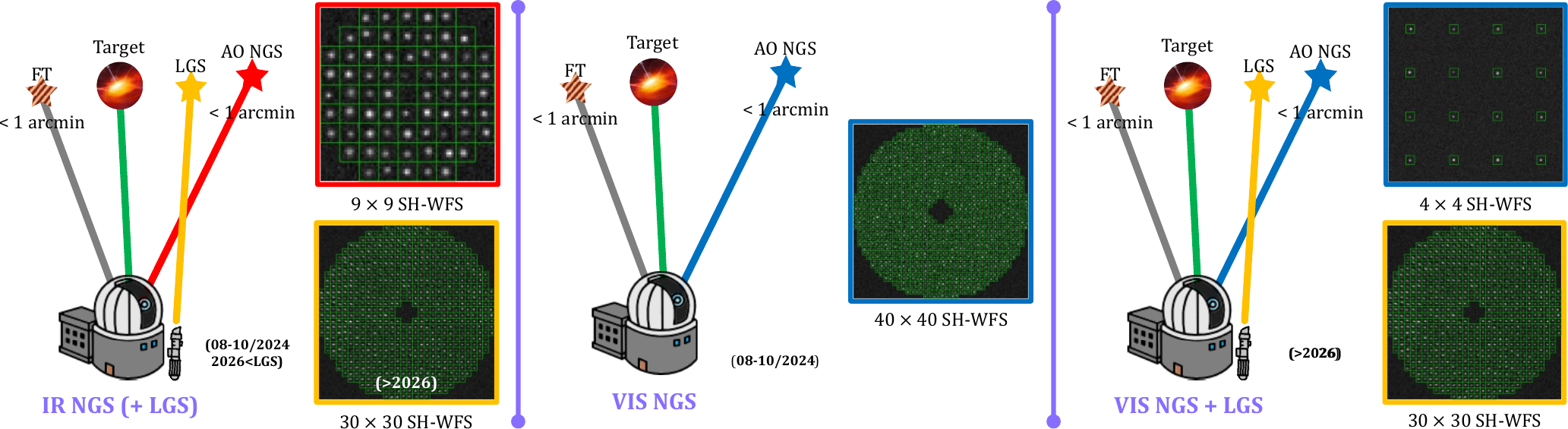}
               
        \caption{\label{fig:GPAO_modes} Upcoming new adaptive optics modes of GRAVITY+. Prior the arrival of the laser guide stars planned in 2026, the \X{9} IR mode will be used as a pure NGS mode, and with a \X{30} LGS afterwards. Two visible modes will be proposed: a pure \X{40} NGS modes for bright targets and a \X{2} TT-NGS supported by the \X{30} LGS for faint targets.
        }
\end{figure}

The GRAVITY+ adaptive optics\cite{LeBouquin:23_GPAO_design} (GPAO) features a \X{43} DM with about 1200 actuators within the \SI{100}{\milli\meter} pupil and different Shack-Hartmann wavefront sensors \cite{Shack:71_SHWFS} (SH-WFSs) depending of the considered mode as shown in \reffig{fig:GPAO_modes}: (i) the current low order (LO) \X{9} infrared (IR) natural guide star (NGS) WFS of the GRAVITY Coudé Infrared Adaptive Optics \cite{Kendrew:12_CIAO}~; (ii) this IR WFS will be coupled with a high order (HO) \X{30} laser guide star (LGS) WFS after the installation of the lasers on the different UTs in 2026~; (iii) a HO \X{40} visible (VIS) NGS WFS mode~; and (iv) a LO \X{2} VIS NGS WFS combined with the HO \X{30} LGS WFS. The choice of the mode depends on a combination of the performances achieved by the fringe tracker (FT) of GRAVITY+ and those on the scientific target.

\begin{figure}[ht!] 
        \centering
        
        \newcommand{\PathFig}{Fig_ASPRO_GPAO_modes/}
        
        \includegraphics[width=0.75\linewidth]{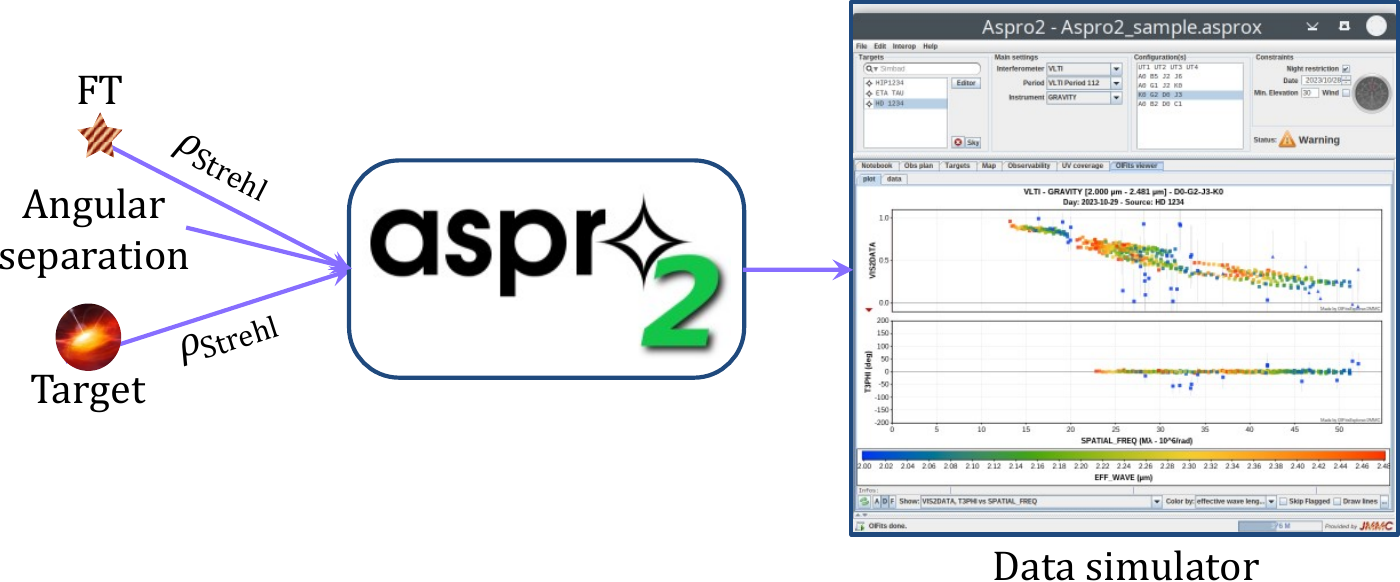}
               
        \caption{\label{fig:ASPRO_2} Basic functioning of the \aspro software.
        }
\end{figure}

To plan the observations and choose the best observing strategy, the users can rely on the 
Astronomical Software to PRepare Observations (\aspro) tool provided by the Jean-Marie Mariotti Center (JMMC). As schematized in \reffig{fig:ASPRO_2}, the Strehl ratios\cite{Roddier:81} on the different sources must be provided to simulate the instrumental performances and predict the science target observability with sufficient signal over noise ratio (S/R). This tool must be updated with the arrival of the different modes of GPAO. Among its objectives, this updated software must allow to choose the best star to feed the FT and the NGS WFS. In addition, it will be used to choose where to place the LGS among the FT, the NGS and the scientific target to maximize the S/R of the object observability curves: (i) on top of the FT to optimize the FT signal?~; (ii) on top of the target to maximize its injection in the GRAVITY+ fibers~; or (iii) on top of the NGS to maximize the AO loop performances? And among the requirements, these new functionalities must be (i) fast to be able to rank a huge number of stars and VLTI configurations for the user to choose among different solutions and (ii) portable to different platforms and easily integrable to the existing softwares.

To answer these needs, we aimed to develop Maréchal approximations\cite{Maréchal:47, Born:80, Ross:09} that parametrize the different modes of GPAO. Such an approximation gives the Strehl ratios as follows\cite{Roddier:81}:
\begin{equation}
	\label{eq:Strehl}
	\strehl = e^{-\sigTwo{tot}}
	\,,
\end{equation}
where $\sigTwo{tot}$ is the total source of wavefront errors in the telescope pupil. As detailed in \citet{Conan:94_PHD}, this Strehl ratio only focuses on its coherent part, which holds the coherent energy, the meaningful observable in the context of interferometry.

In the following, we first introduce in \refsec{sec:param} the different parameters playing a role in the GPAO modes. Then, in \refsec{sec:NGS}, we present the Maréchal approximation for the NGS modes of GPAO, both VIS and IR, and their calibration using TIPTOP, see \citet{Neichel:20_TIPTOP}, a toolkit to predict long term exposure point spread function (PSF) via an analytical model based on a spatial power spectrum density (PSD) approach of the turbulence residuals in the pupil after AO correction\cite{Jolissaint:06_PSD, Jolissaint:10_PSD}. The Maréchal approximation for the LGS modes is finally described in \refsec{sec:LGS}. A recent update of TIPTOP, see \citet{Agapito:23_AO4ELT_TIPTOP_cone}, included the cone effect produced by laser guide stars, making the tool also suitable for the calibration of these more complex modes.

\section{Actors of the play}
\label{sec:param}

The performances of an AO system depend on many key parameters. Some are linked with the design of the AO system itself, which are fixed once for all once the system is built. Others depend on the observation conditions, such as the turbulence statistics or the observed targets. The parameters used in the different GPAO models are listed in \reftabs{tab:parameter_atm}{tab:parameter_AO}.

\begin{table}[th!] 
    \caption{\label{tab:parameter_atm} Notation and description of the atmosphere and source parameters.}
    \centering
    \begin{tabular}{cc}
    \hline
    \hline
    $s\in\Brace{\text{sci}, \text{ngs}, \text{lgs}}$ & Source: science, natural guide star, laser guide star
    \\
    $\lambdaTag{0}$ & Reference wavelength for the atmosphere
    \\
    $\lambdaTag{$s$}$ & Wavelength of the source $s$
    \\
    $\rFried{$s$}$ & Fried parameter at $\lambdaTag{$s$}$ of the turbulence layer
    \\
    $\hWind$ & Altitude of the turbulence layer
    \\
    $\hTag{lgs}$ & Altitude of the laser guide star
    \\
    $\vWind$ & Velocity of the turbulence layer
    \\
    $\thetaTag{$s$,$s^\prime$}$ & Angular separation between the sources $s$ and $s^\prime$
    \\
    $\secant = \text{sec}\,\zeta = 1/\cos\,\zeta$ & Secant of the zenith angle $\zeta$
    \\
    $\fluxTag{$s$}$ & \makecell{Flux (ph/m$^2$/s) of the source $s$ \\ (accounting for the instrument transmission)}
    \\
    \hline
    \end{tabular}
\end{table}

\begin{table}[th!] 
    \caption{\label{tab:parameter_AO} Notation and description of the parameters of the AO system.}
    \centering
    \begin{tabular}{cc}
    \hline
    \hline
    $g$ & AO loop gain
    \\
    $f$ & Frequency of the loop gain
    \\
    $\dtel$ & Telescope diameter
    \\
    $\nTag{modes}$ & Number of modes controlled by the AO system
    \\
    $\dact\triangleq \dtel / \Paren{2\sqrt{\nTag{modes}/\pi}-1}$ & Equivalent inter-actuator distance
    \\
    $\nTag{wfs}$ & Number of lenslets across the SH-WFS diameter
    \\
    $\dwfs = \dtel/\nTag{wfs}$ & Diameter of a SH-WFS subaperture
    \\
    $\sig{pix}$ & Pixel readout noise
    \\
    $\pixscale$ & Pixel scale of the SH-WFS
    \\
    $\nTag{pix}$ & Number of pixels across a SH-WFS box
    \\
    $\fwhmTag{$s$}$ & Full width at half maximum (FWHM) of the source $s$
    \\
    $\nTag{ph,$s$}=\fluxTag{$s$}\frac{\dwfs^2}{f}$ & \makecell{Number of photons in a SH-WFS box \\ for the source $s$ during one short exposure}
    \\
    \hline
    \end{tabular}
\end{table}

We notice that the science $s=\Tag{sci}$ can either be the scientific target or the FT star depending on the situation. We also emphasize that the NGS can be used to close the HO loop or close the LO loop, mainly for tip-tilt control, when using a LGS. As a reminder\cite{Roddier:81}, we recall here the scaling equation for the Fried parameter,
\begin{equation}
	\label{eq:rFried}
	\rFried{s} = \rFried{0}\Paren{\frac{\lambdaTag{s}}{\lambdaTag{0}}}^{6/5}
	\,.
\end{equation}

\section{Natural guide star modes}
\label{sec:NGS}

\subsection{Maréchal Approximation}

They are four main errors driving an AO system\cite{Rigaut:98_SH_error}: (i) the fitting error, limited by the shapes that the DM can take to compensate the turbulence~; (ii) the aliasing error, limited by the bandwidth of the WFS that wraps high spatial frequencies of the turbulence to low order errors~; (iii) the noise error, with the camera readout noise and photon noise propagating through the loop~; (iv) the servo-lag error, that comes from the fact that an AO loop is always `late', with a DM correction applied with some delay
because of the WFS exposure time and RTC computation loads. The fitting and aliasing errors have similar properties and dependencies. In the following, they will be grouped in a single term: the geometrical error.

On top of the core errors listed above, additional terms must be added when shifting from single conjugated AO to more complex configurations. In particular, for the NGS modes, the isoplanetism error plays a critical role when the scientific target is not the source on which the HO AO loop is closed, due to the turbulence decorrelation in function of their angular separation.

Adding all the mentioned errors, the total variance in \refeq{eq:Strehl} for the NGS modes writes as follows:
\begin{equation}
	\label{eq:NGS}
	\sigTwo{ngs} = \sigTwo{0} + \sigTwo{geom} + \sigTwo{lag} + \sigTwo{ph} + \sigTwo{ron} + \sigTwo{iso}
	\,,
\end{equation}
where $\sigTwo{0}$ represents the absolute limit achievable when all the other terms are canceled. This offset may include other errors not accounted for by our model, such as static residuals errors, vibrations, SH-WFS aliasing limit and so on. The dependency of the other different terms of \refeq{eq:NGS} with respect to the parameters introduced in \refsec{sec:param} is as follows:
\begin{align}
	\label{eq:NGS_geom}
	\text{geometry} & \rightarrow \alphaTag{geom} \cdot \Paren{\frac{\dact}{\secant^{-3/5}\rFried{sci}}}^{5/3}
	\,,
	\\
	\label{eq:NGS_lag}
	\text{servo-lag} & \rightarrow \alphaTag{lag} \cdot \Paren{\frac{\vWind}{\secant^{-3/5}\rFried{sci} f g}}^{\betaTag{lag}}
	\,,
	\\
	\label{eq:NGS_ph}
	\text{photon noise} & \rightarrow \alphaTag{ph} \cdot \Paren {\frac{\fwhmTag{sci}}{\lambdaTag{sci}/\dTag{wfs}}}^2 \cdot 2 \nTag{ph,ngs} \cdot 1/\nTag{ph,ngs}^2 \cdot \frac{g}{1-g}
	\,,
	\\
	\label{eq:NGS_noise}
	\text{readout noise} & \rightarrow \alphaTag{ron} \cdot \pixscale^2 \nTag{pix}^4\sigTwo{pix} \cdot 1/\nTag{ph,ngs}^2 \cdot \frac{g}{1-g}
	\,,
	\\
	\label{eq:NGS_iso}
	\text{isoplanetism} & \rightarrow \alphaTag{iso} \cdot  \Paren{\frac{\thetaTag{sci,ngs}\secant\hWind}{\secant^{-3/5}\rFried{sci}}}^{\betaTag{iso}}
	\,.
\end{align}

The $1/\secant^{-3/5}$ dependency in \refeqss{eq:NGS_geom}{eq:NGS_lag}{eq:NGS_iso} is here to scale the Fried parameter with the airmass. The $\beta$ coefficients in \refeqs{eq:NGS_lag}{eq:NGS_iso} are here to permit a deviation from the classical $5/3$ of a Kolmogorov statistic in the presence of an external scale for the structure function of the turbulence. The gain dependency in \refeqs{eq:NGS_ph}{eq:NGS_noise} is discussed in \citet{Vidal:14_gain}. In \refeq{eq:NGS_ph}, (i) the first term is the ratio of the spot positioning uncertainty scaled to the science wavelength, (ii) the second term is the shot noise variance with a factor 2 accounting for the excess of electron-multiplied sensors, and (iii) the third term is the signal variance, increasing with the squared number of photons. In \refeq{eq:NGS_noise}, (i) the first term is the readout noise variance, which also depends on the centroiding method (hidden in the $\alphaTag{ron}$ coefficient), and (ii) the second term is the signal variance, increasing with the squared number of photons.

\subsection{Confrontation with TIPTOP}

The different scaling factors $\alpha$ and $\beta$ in Eqs.~(\ref{eq:NGS_geom}-\ref{eq:NGS_iso}) were calibrated using the TIPTOP toolbox. A default configuration for the VIS and IR-modes was used, as listed in \reftab{tab:default_NGS}, using an external scale of the turbulence structure function of \SI{22}{\meter}. The default NGS magnitude is 8 with a conversion factor for a 0-magnitude star of $2.63\times10^{10}\,$\SI{}{ph\per\second\per\meter\squared} for the VIS modes and $1.66\times10^{9}\,$\SI{}{ph\per\second\per\meter\squared} for the IR modes.

\begin{table}[th!] 
    \caption{\label{tab:default_NGS} Default configuration of the VIS and IR-NGS modes.}
    \centering
    \begin{tabular}{cc}
    \hline
    \hline
    Parameter & VIS/IR-NGS
    \\
    \hline
    $\dtel$ & \SI{8}{\meter}
    \\
    $\zeta$ & \SI{0}{\degree} (pointing at zenith, $\secant = 1$)
    \\
    Seeing & 1''
    \\
    $\hWind$ & \SI{10}{\kilo\meter}
    \\
    $\vWind$ & \SI{25}{\meter\per\second}
    \\
    $\thetaTag{sci,ngs}$ & \SI{0}{\degree} (on-axis)
    \\
    $\lambdaTag{0}$ & \SI{500}{\nano\meter}
    \\
    $\lambdaTag{sci}$ & \SI{2.2}{\micro\meter}
    \\
    $\lambdaTag{ngs}$ & \SI{750}{\nano\meter} / \SI{2.2}{\micro\meter}
    \\
    $\nTag{modes}$ & 800 $\Paren{\dact = \SI{0.28}{\meter}}$ / 44 $\Paren{\dact = \SI{1.23}{\meter}}$
    \\
    $\nTag{wfs}$ & 40 / 9
    \\
    $\sig{pix}$ & $0.2\,e/\text{frame}/\text{pix}$
    \\
    $\pixscale$ & 0.42'' / 0.51''
    \\
    $\nTag{pix}$ & 6 / 8
    \\
    $g$ & 0.5
    \\
    $f$ & \SI{1000}{\hertz} / \SI{500}{\hertz}
    \\
    $\nTag{ph,ngs}$ & 200 / 250 (8-magnitude NGS)
    \\
    \hline
    \end{tabular}
\end{table}

Each error is fitted one by one, by varying one of its characteristic coefficients: (i) the actuator pitch $\dact$ for the geometric error (equivalent to the number of modes), see \refsubfig{fig:TIPTOP_fit}{a}, (ii) the wind speed $\vWind$ for the servo-lag error, see \refsubfig{fig:TIPTOP_fit}{b}, (iii) the number of photons $\nTag{ph}$ for the photon noise error, see \refsubfig{fig:TIPTOP_fit}{c}, (iv) the readout noise $\sig{pix}$ for the readout noise error, see \refsubfig{fig:TIPTOP_fit}{d}, and (v) the angular separation between the NGS and the science target $\thetaTag{ngs,sci}$ for the anisoplanetism error, see \refsubfig{fig:TIPTOP_fit}{e}. To assess the absolute Strehl limit $\strehl^{0}=e^{-\sigTwo{0}}$ achievable with all error tending towards zero, the geometric error was fitted as described in the figure caption by canceling all other error terms. In \refsubfig{fig:TIPTOP_fit}{c}, the photon noise was also decoupled from the readout noise.

\begin{figure}[t!] 
        \centering
        
        \newcommand{\PathFig}{Fig_ASPRO_TIPTOP_NGS/}

        \newcommand{\fontTxt}[1]{\textbf{\small $\quad\quad$ #1}}        
        
        \newcommand{\LineRatio}{0.95}
        
        \newcommand{\subfigColor}{black}
        
        \newcommand{\figOne}{\PathFig TIPTOP_vs_geom.png}
        \newcommand{\figTwo}{\PathFig TIPTOP_vs_lag.png}
        \newcommand{\figThree}{\PathFig TIPTOP_vs_ph.png}
        
        \sbox1{\includegraphics{\figOne}}
        \sbox2{\includegraphics{\figTwo}}
        \sbox3{\includegraphics{\figThree}}
        
        \newcommand{\ColumnWidth}[1]
                {\dimexpr \LineRatio \linewidth * \AspectRatio{#1} / (\AspectRatio{1} + \AspectRatio{2} + \AspectRatio{3}) \relax
                }
        \newcommand{\ColumnGap}{\hspace {\dimexpr \linewidth /4 - \LineRatio\linewidth /4 }}

        \begin{tabular}{
                @{\ColumnGap}
                M{\ColumnWidth{1}}
                @{\ColumnGap}
                M{\ColumnWidth{2}}
                @{\ColumnGap}
                M{\ColumnWidth{3}}
                @{\ColumnGap}
                }
                
                \fontTxt{Geometric error}
                &
                \fontTxt{Servo-lag error}
                &
                \fontTxt{Photon noise error}
                \\  
                \subfigimg[width=\linewidth,pos=ul,font=\fontfig{\subfigColor}]{$\!\!\!\!$(a)}{0.0}{\figOne} &
                \subfigimg[width=\linewidth,pos=ul,font=\fontfig{\subfigColor}]{$\!\!\!\!$(b)}{0.0}{\figTwo} &
                \subfigimg[width=\linewidth,pos=ul,font=\fontfig{\subfigColor}]{$\!\!\!\!$(c)}{0.0}{\figThree}              
        \end{tabular}

        \renewcommand{\LineRatio}{0.65}        
        
        \renewcommand{\figOne}{\PathFig TIPTOP_vs_ron.png}
        \renewcommand{\figTwo}{\PathFig TIPTOP_vs_iso.png}
        
        \sbox1{\includegraphics{\figOne}}
        \sbox2{\includegraphics{\figTwo}}
        
        \renewcommand{\ColumnWidth}[1]
                {\dimexpr \LineRatio \linewidth * \AspectRatio{#1} / (\AspectRatio{1} + \AspectRatio{2}) \relax
                }
        \renewcommand{\ColumnGap}{\hspace {\dimexpr \linewidth /3 - \LineRatio\linewidth /3 }}

        \begin{tabular}{
                @{\ColumnGap}
                M{\ColumnWidth{1}}
                @{\ColumnGap}
                M{\ColumnWidth{2}}
                @{\ColumnGap}
                }
                
                \fontTxt{Readout noise}
                &
                \fontTxt{Anisoplanetism error}
                \\  
                \subfigimg[width=\linewidth,pos=ul,font=\fontfig{\subfigColor}]{$\!\!\!\!$(d)}{0.0}{\figOne} &
                \subfigimg[width=\linewidth,pos=ul,font=\fontfig{\subfigColor}]{$\!\!\!\!$(e)}{0.0}{\figTwo}             
        \end{tabular}   
               
        \caption{\label{fig:TIPTOP_fit} Fitting the coefficients of \reftab{tab:fitted_coef_NGS} with TIPTOP by varying parameters associated with each sources of error and using the default configuration of \reftab{tab:default_NGS}. \refpan{a}: Default configuration changed as follows: $\nTag{ph,ngs}=10000$, $\vWind = \SI{0.01}{\meter\per\second}$ and $\sig{pix} = 0\,e/\text{frame}/\text{pix}$. \refpan{c}: Default configuration changed as follows: $\sig{pix} = 0\,e/\text{frame}/\text{pix}$. \refpan{e}: The wind is orthogonal to the separation angle.}
\end{figure}

The different fitted coefficients are given in \reftab{tab:fitted_coef_NGS}. The lower value of $\strehl^{0}$ for the \X{30} IR-NGS can be imputable to the strong aliasing induced by the strong under-sampling of the SH-WFS relative to the high resolution DM. The $\alphaTag{geom}$ are in the classical range\cite{Rigaut:98_SH_error} of $\Brack{0.2,0.3}$. All the $\beta$ terms are consistent, a bit higher than the classical $5/3\simeq 1.67$ Kolmogorov approximation.

\begin{figure}[hb!] 
        \centering
        
        \newcommand{\PathFig}{Fig_ASPRO_TIPTOP_NGS/}

        \newcommand{\fontTxt}[1]{\textbf{\small $\quad\quad$ #1}}        
        
        \newcommand{\LineRatio}{0.66}
        
        \newcommand{\subfigColor}{black}
        
        \newcommand{\figOne}{\PathFig Cross_vs_wavelength.png}
        \newcommand{\figTwo}{\PathFig Cross_vs_frequency.png}
        
        \sbox1{\includegraphics{\figOne}}
        \sbox2{\includegraphics{\figTwo}}
        
        \newcommand{\ColumnWidth}[1]
                {\dimexpr \LineRatio \linewidth * \AspectRatio{#1} / (\AspectRatio{1} + \AspectRatio{2}) \relax
                }
        \newcommand{\ColumnGap}{\hspace {\dimexpr \linewidth /3 - \LineRatio\linewidth /3 }}

        \begin{tabular}{
                @{\ColumnGap}
                M{\ColumnWidth{1}}
                @{\ColumnGap}
                M{\ColumnWidth{2}}
                @{\ColumnGap}
                }
                
                \fontTxt{\refeq{eq:NGS} vs $\lambdaTag{sci}$}
                &
                \fontTxt{\refeq{eq:NGS} vs $f$}
                \\  
                \subfigimg[width=\linewidth,pos=ul,font=\fontfig{\subfigColor}]{$\!\!\!\!$(a)}{0.0}{\figOne} &
                \subfigimg[width=\linewidth,pos=ul,font=\fontfig{\subfigColor}]{$\!\!\!\!$(b)}{0.0}{\figTwo}              
        \end{tabular}

        \renewcommand{\figOne}{\PathFig Cross_vs_gain_1.png}
        \renewcommand{\figTwo}{\PathFig Cross_vs_gain_2.png}
        \sbox1{\includegraphics{\figOne}}
        \sbox2{\includegraphics{\figTwo}}

        \begin{tabular}{
                @{\ColumnGap}
                M{\ColumnWidth{1}}
                @{\ColumnGap}
                M{\ColumnWidth{2}}
                @{\ColumnGap}
                }
                
                \fontTxt{\refeq{eq:NGS} vs $g$ (servo-lag)}
                &
                \fontTxt{\refeq{eq:NGS} vs $g$ (photon noise)}
                \\  
                \subfigimg[width=\linewidth,pos=ul,font=\fontfig{\subfigColor}]{$\!\!\!\!$(c)}{0.0}{\figOne} &
                \subfigimg[width=\linewidth,pos=ul,font=\fontfig{\subfigColor}]{$\!\!\!\!$(d)}{0.0}{\figTwo}              
        \end{tabular}

        \renewcommand{\LineRatio}{0.95}        
        
        \renewcommand{\figOne}{\PathFig Cross_vs_zenith_lag.png}
        \renewcommand{\figTwo}{\PathFig Cross_vs_zenith_aniso.png}
        \newcommand{\figThree}{\PathFig Cross_vs_layer.png}
        
        \sbox1{\includegraphics{\figOne}}
        \sbox2{\includegraphics{\figTwo}}
        \sbox3{\includegraphics{\figThree}}
        
        \renewcommand{\ColumnWidth}[1]
                {\dimexpr \LineRatio \linewidth * \AspectRatio{#1} / (\AspectRatio{1} + \AspectRatio{2} + \AspectRatio{3}) \relax
                }
        \renewcommand{\ColumnGap}{\hspace {\dimexpr \linewidth /4 - \LineRatio\linewidth /4 }}

        \begin{tabular}{
                @{\ColumnGap}
                M{\ColumnWidth{1}}
                @{\ColumnGap}
                M{\ColumnWidth{2}}
                @{\ColumnGap}
                M{\ColumnWidth{3}}
                @{\ColumnGap}
                }
                
                \fontTxt{\refeq{eq:NGS} vs $\zeta$ (servo-lag)}
                &
                \fontTxt{\refeq{eq:NGS} vs $\zeta$ (isoplanetism)}
                &
                \fontTxt{\refeq{eq:NGS} vs multi-layers}
                \\  
                \subfigimg[width=\linewidth,pos=ul,font=\fontfig{\subfigColor}]{$\!\!\!\!$(e)}{0.0}{\figOne} &
                \subfigimg[width=\linewidth,pos=ul,font=\fontfig{\subfigColor}]{$\!\!\!\!$(f)}{0.0}{\figTwo} &
                \subfigimg[width=\linewidth,pos=ul,font=\fontfig{\subfigColor}]{$\!\!\!\!$(g)}{0.0}{\figThree}             
        \end{tabular}   
               
        \caption{\label{fig:TIPTOP_cross} Validating \refeq{eq:NGS} and the fitted coefficients of \reftab{tab:fitted_coef_NGS} on parameters cross-coupling the sources of errors using the default configuration of \reftab{tab:default_NGS}. For information, the long exposure PSF predicted by TIPTOP are given in some panels (log scale). \refpan{c}: Default configuration changed as follows: $\nTag{ph,ngs}=200$ and $\vWind = \SI{35}{\meter\per\second}$. \refpan{d}: Default configuration changed as follows: $\nTag{ph,ngs}=10$ and $\vWind = \SI{0.1}{\meter\per\second}$.\refpan{e}: Default configuration changed as follows: $\nTag{ph,ngs}=10^5$, $\vWind = \SI{75}{\meter\per\second}$, $\sig{pix} = 0\,e/\text{frame}/\text{pix}$. \refpan{f}: Default configuration changed as follows: $\nTag{ph,ngs}=10^5$, $\vWind = 10^{-4}\,\SI{}{\meter\per\second}$, $\sig{pix} = 0\,e/\text{frame}/\text{pix}$, $\thetaTag{sci,ngs}=2''$. \refpans{f,g}: Wind direction orthogonal to the source separation.}
\end{figure}

To further validate the Maréchal approximation, \refeq{eq:NGS} and the fitted coefficients of \reftab{tab:fitted_coef_NGS} are tested by varying other parameters, as shown in \reffig{fig:TIPTOP_cross}. These parameters induce cross-coupling of different error terms and permit to check that no physical dependency was forgotten in the model. \refsubfigsfull{fig:TIPTOP_cross}{a,b} show a very good agreement with the wavelength~$\lambdaTag{sci}$ of the target and the AO loop frequency~$f$. Two regimes are tested for the AO loop gain $g$, see \refsubfigs{fig:TIPTOP_cross}{c,d}: a situation where the servo-lag is the limiting factor and another one where the photon and readout noises are the limiting factors. The global trends are correct, except for unrealistic gains below 0.2 or above 0.7. Two different regimes were also tested concerning the impact of the airmass, see \refsubfigs{fig:TIPTOP_cross}{e,f}: a limitation by the servo-lag error and by the anisoplanetism error. The agreement between TIPTOP and our simplified model is very good on the full Strehl dynamics. For information, some long exposure PSFs predicted by TIPTOP are shown in insets.

\begin{table}[h!] 
    \caption{\label{tab:fitted_coef_NGS} Coefficients of the Maréchal approximation of \refeq{eq:NGS} fitted with TIPTOP.}
    \centering
    \begin{tabular}{ccc}
    \hline
    \hline
    Parameter & \X{40} VIS-NGS & \X{30} IR-NGS
    \\
    \hline
    $\strehl^{0}$ & \percent{99} & \percent{86} 
    \\
    $\alphaTag{geom}$ & 0.27 & 0.24
    \\
    $\alphaTag{lag}\;/\;\betaTag{lag}$ & 8.48 / 2.16 & 2.08 / 2.10
    \\
    $\alphaTag{ph}$ & 12.0 & 15.2  
    \\
    $\alphaTag{ron}$ & 0.52 & 1.65 
    \\
    $\alphaTag{iso}\;/\;\betaTag{iso}$ & 4.34 / 1.86 & 1.75 / 1.97
    \\
    \hline
    \end{tabular}
\end{table}

Finally, the Maréchal approximation was tested against a multi-layered atmosphere. Two layers are included in the model: a first layer at $h_l=\SI{7.5}{\kilo\meter}$ with a wind speed of $v_l=\SI{35}{\meter\per\second}$ and a second layer at $h_l=\SI{12.5}{\kilo\meter}$ with a wind speed of $v_l=\SI{100}{\meter\per\second}$, with a mixing coefficient $C_n^2$. As seen above, the Maréchal approximation only accepts one coefficient corresponding to the mean parameter of the turbulence. It is computed using the classical formula for the parameter $\gamma_l$\cite{Roddier:81}:
\begin{equation}
	\label{eq:Cn2}
	\gamma_0 = \Paren{\sum_{l=1}^{n_\Tag{layer}}{C_{n,l}^2\cdot \gamma_l^{5/3}}}^{3/5}
	\,.
\end{equation}
TIPTOP  can take multi-layered atmospheres as an input. \refsubfigfull{fig:TIPTOP_cross}{g} shows the Strehl ratio predicted by TIPTOP according to $C_{n,2}^2$. The matching with the Maréchal approximation is globally very good, both in terms of trend and values. This comforts the fact that characteristics parameters of the turbulence, measured by the atmosphere monitors on site, can be used to feed the Maréchal approximation.

\section{Towards the laser guide star modes}
\label{sec:LGS}

Deriving the Maréchal approximation for LGS modes is more complex. Indeed, the wavefront control is split in two sub-loops: (i) the low order loop closed on a faint NGS mainly for TT correction, and (ii) the high order loop on the LGS. These two loops suffer from the same errors mentioned in \refsec{sec:NGS}, plus the additional error of the partial correction of the HO by the LGS due to the cone effect\cite{Agapito:23_AO4ELT_TIPTOP_cone}. This leads to the following error budget:
\begin{equation}
	\label{eq:LGS}
	\sigTwo{lgs} = \sigTwo{0} + \sigTwo{geom} + \sigTwo{lag,HO} + \sigTwo{lag,LO} + \sigTwo{ph,HO} + \sigTwo{ph,LO} + \sigTwo{ron,HO} + \sigTwo{ron,LO} + \sigTwo{iso,HO} + \sigTwo{iso,LO} + \sigTwo{cone,sci}
	\,.
\end{equation}
The dependency of the different terms of \refeq{eq:LGS} with respect to the parameters introduced in \refsec{sec:param} for the LO and HO components is as follows:
\begin{align}
	\label{eq:LGS_geom}
	\text{geometry} & \rightarrow \alphaTag{geom} \cdot \Paren{\frac{\dact}{\secant^{-3/5}\rFried{sci}}}^{5/3}
	\,,
	\\
	\label{eq:LGS_lag}
	\text{servo-lag (HO/LO)} & \rightarrow \alphaTag{lag,HO/LO} \cdot \Paren{\frac{\vWind}{\secant^{-3/5}\rFried{sci} f_\Tag{HO/LO} g_\Tag{HO/LO}}}^{\betaTag{lag,HO/LO}}
	\,,
	\\
	\label{eq:LGS_ph}
	\text{photon noise (HO/LO)} & \rightarrow \alphaTag{ph,HO/LO} \cdot  \Paren {\frac{\fwhmTag{lgs/ngs}}{\lambdaTag{sci}/\dTag{wfs,HO/LO}}}^2 \cdot 2 \nTag{ph,lgs/ngs} \cdot 1/\nTag{ph,lgs/ngs}^2 \cdot \frac{g_\Tag{HO/LO}}{1-g_\Tag{HO/LO}} 
	\,,
	\\
	\label{eq:LGS_noise}
	\text{readout noise (HO/LO)} & \rightarrow \alphaTag{ron,HO/LO} \cdot  \pixscaleTag{HO/LO}^2 \nTag{pix,HO/LO}^4\sigTwo{pix,HO/LO} \cdot 1/\nTag{ph,lgs/ngs}^2 \cdot \frac{g_\Tag{HO/LO}}{1-g_\Tag{HO/LO}}
	\,,
	\\
	\label{eq:LGS_iso_HO}
	\text{isoplanetism (HO)} & \rightarrow \alphaTag{iso,HO} \cdot  \Paren{\frac{\thetaTag{sci,lgs}\secant\hWind}{\secant^{-3/5}\rFried{sci}}}^{\betaTag{iso,HO}}
	\,,
	\\
	\label{eq:LGS_iso_LO}
	\text{isokinetism (LO)} & \rightarrow \alphaTag{iso,LO}  \cdot \Paren{\frac{\thetaTag{sci,ngs}\secant\hWind}{\secant^{-3/5}\rFried{sci}}}^{\betaTag{iso,LO}}
	\,,
	\\
	\label{eq:LGS_cone}
	\text{cone effect} & \rightarrow \alphaTag{cone,sci} \cdot  \Paren{\frac{\dtel}{\secant^{-3/5}\rFried{sci}}\frac{\hWind}{\hTag{lgs}}}^{\betaTag{cone,sci}}
	\,.
\end{align}
Most of the terms are similar to the NGS terms, but duplicated on the two loops. \refeqfull{eq:LGS_iso_LO} represents the isokinetism error, mainly applying to the TT variance in the wavefront residuals.
In a first approximation, the FWHMs of \refeq{eq:LGS_ph} are given as follows:
\begin{align}
	\label{eq:FWHM_lgs}
	\fwhmTag{lgs} \simeq & 1'' \text{ (LGS spot size)}
	\,,
	\\
	\label{eq:FWHM_ngs}
	\fwhmTag{ngs}^2 \simeq & \strehl^\Tag{LO}\Paren{\frac{\lambdaTag{lgs}}{\dTag{wfs,LO}}}^2 + \Paren{1-\strehl^\Tag{LO}}\Paren{\frac{\lambdaTag{lgs}}{\rFried{lgs}}}^2
	\,,
\end{align}
where $\strehl^\Tag{LO} = e^{-\sigTwo{LO}}$ is the Strehl ratio on the LO WFS. \refeqfull{eq:FWHM_ngs} emphasizes the trade-off between a seeing limited and a diffraction limited spot in the LO WFS, depending on the performances of the HO loop. Thus, $\sigTwo{LO}$ is given by:
\begin{equation}
	\label{eq:LGS_LO}
	\sigTwo{LO} = \sigTwo{geom} + \sigTwo{lag,HO}  + \sigTwo{ph,HO} + \sigTwo{ron,HO} + \sigTwo{iso,HO} + \sigTwo{iso} + \sigTwo{cone,lgs}
	\,,
\end{equation}
with:
\begin{align}
	\label{eq:LO_geom}
	\text{geometry} & \rightarrow \alphaTag{geom} \cdot \Paren{\frac{\dact}{\secant^{-3/5}\rFried{ngs}}}^{5/3}
	\,,
	\\
	\label{eq:LO_lag}
	\text{servo-lag} & \rightarrow \alphaTag{lag} \cdot \Paren{\frac{\vWind}{\secant^{-3/5}\rFried{ngs} f_\Tag{HO} g_\Tag{HO}}}^{\betaTag{lag,HO}}
	\,,
	\\
	\label{eq:LO_ph}
	\text{photon noise} & \rightarrow \alphaTag{ph} \cdot \Paren {\frac{\fwhmTag{lgs}}{\lambdaTag{ngs}/\dTag{wfs,HO}}}^2 \cdot 2 \nTag{ph,lgs} \cdot 1/\nTag{ph,lgs}^2 \cdot \frac{g_\Tag{HO}}{1-g_\Tag{HO}}
	\,,
	\\
	\label{eq:LO_noise}
	\text{readout noise} & \rightarrow \alphaTag{ron} \cdot \pixscaleTag{HO}^2 \nTag{pix,HO}^4\sigTwo{pix,HO} \cdot 1/\nTag{ph,lgs}^2 \cdot \frac{g_\Tag{HO}}{1-g_\Tag{HO}}
	\,,
	\\
	\label{eq:LO_iso_HO}
	\text{isoplanetism} & \rightarrow \alphaTag{iso} \cdot \Paren{\frac{\thetaTag{lgs,ngs}\secant\hWind}{\secant^{-3/5}\rFried{ngs}}}^{\betaTag{iso}}
	\,,
	\\
	\label{eq:LO_cone}
	\text{cone effect} & \rightarrow \alphaTag{cone,LO} \cdot \Paren{\frac{\dTag{wfs,LO}}{\secant^{-3/5}\rFried{ngs}}\frac{\hWind}{\hTag{lgs}}}^{\betaTag{cone,LO}}
	\,.
\end{align}
All the LO terms disappear in this budget. Indeed, the formula of \refeq{eq:LGS_LO} must gives the statistic of the instantaneous Strehl ratio in the LO WFS. We consequently work under the assumptions that the LO loop freezes the low orders and that the high orders are averaged to get a statistical meaning of the Strehl in the LO loop. The coefficients in Eqs.~(\ref{eq:LO_geom}-\ref{eq:LO_cone}) should share the same values as the one fitted for Eqs.~(\ref{eq:LGS_geom}-\ref{eq:LGS_cone}). But these coefficients should be different than the ones fitted in \refsec{sec:NGS} due to the different geometry of the LGS loop and the associated cone effect.

There are consequently lots of cross-couplings among the different terms. A fitting strategy is yet to be defined with TIPTOP for these more complex LGS modes. As well as future refinements of the equations in case of strong discrepancies. This study was not achieved by the time of the conference and will be the subject of a future communication.

\section{Conclusions and perspectives}

In this work, we introduced a simplified model of the different GPAO modes based on Maréchal approximations. We first focused on the NGS modes (VIS and IR). They have been parametrized by trying to grasp the main physical processes impacting the Strehl ratio, driven by key parameters describing the sources and the atmosphere on one side, and the AO system on the other side.

The LGS modes are more complicated due to the presence of a dual channel for HO and LO correction, multiplying the sources of error that impact the Strehl ratio, as well as cross-terms. A formulation has been proposed but needs to be further validated to check the dependencies and its completeness. The commissioning the lasers and the associated modes is not planned before 2026. This lets some time to refine the equations and prepare the observations.

At first, these models aimed to be validated and calibrated with TIPTOP simulations. This has already been achieved with the NGS modes which have then already been integrated in the \texttt{SearchFTT} ranking tool of the JMMC. In a first approximation, they can be used to prepare the instrument commissioning by selecting pertinent sources and associated guide stars to calibrate the GPAO modes and check their performances. This can also been used to prepare a first version of a catalog of scientific sources to be observed with GRAVITY+ associated with their FT star and NGS.

But in the long run, the parameters of the Maréchal approximation, currently fitted with TIPTOP, aim to be refined with more realistic situations. Running complex end-to-end simulations was not considered efficient regarding the amount of work and computation time this would imply. After the GPAO commissioning, the different data acquired in various conditions and instrument settings will rather be used to update the parameters with real data and align the model to the true performances of the instrument.

\acknowledgments 
 
The authors warmly thank Guido Agapito and Benoit Neichel for their availability and reactivity and their great help in using TIPTOP.
\\
This project has received funding from the European Union's Horizon 2020 research and innovation programme under grant agreements No 101004719.
\\
This research has made use of the Jean-Marie Mariotti Center \aspro service, available at \href{http://www.jmmc.fr/aspro}{http://www.jmmc.fr/ aspro}.

\bibliography{2024_SPIE_Berdeu-et-al} 
\bibliographystyle{spiebib} 

\end{document}